\documentstyle[amsfonts,epsfig,12pt]{article}

\def\mypagenumber{1}

\def\myend{\end{document}}

\normalsize

\newcounter{sxn}

\newcounter{axn}

\date{}

\newdimen\mybaselineskip
\newcommand{\beeq}{\begin{equation}}
\newcommand{\eneq}{\end{equation}}
\newcommand{\be}{\begin{eqnarray}}
\newcommand{\ee}{\end{eqnarray}}
\newcommand{\bpic}{\begin{picture}}
\newcommand{\epic}{\end{picture}}

\def\la{\raise.16ex\hbox{$\langle$} \, }
\def\ra{\, \raise.16ex\hbox{$\rangle$} }

\def\psibar{ \psi \kern-.65em\raise.6em\hbox{$-$} }
\def\mbar{ m \kern-.78em\raise.4em\hbox{$-$}\lower.4em\hbox{} }

\def\n@space{\nulldelimiterspace=0pt \mathsurround=0pt }
\def\huge#1{{\hbox{$\left#1\vbox to 20.5pt{}\right.\n@space$}}}

\def\myskip{\noalign{\kern 8pt}}
\def\myeqspace{\noalign{\kern 10pt}}

\def\boxit#1{$\vcenter{\hrule\hbox{\vrule\kern3pt
    \vbox{\kern3pt\hbox{#1}\kern3pt}\kern3pt\vrule}\hrule}$}
\def\bigbox#1{$\vcenter{\hrule\hbox{\vrule\kern5pt
     \vbox{\kern5pt\hbox{#1}\kern5pt}\kern5pt\vrule}\hrule}$}

\def\ignore#1{{}}

\begin{document}
\bibliographystyle{unsrt}
\footskip 1.0cm

\thispagestyle{empty}
\setcounter{page}{\mypagenumber}


\begin{flushright}{
BRX-TH-508\\}

\end{flushright}

\vspace{2.5cm}
\begin{center}
{\LARGE \bf {Partially  Massless Spin-2 Fields in 
String Generated Models }}\\
\vskip 1 cm
{\large{Bayram Tekin  }}\footnote{e-mail:~
tekin@brandeis.edu}\\
\vspace{.5cm}
{\it Department of  Physics, Brandeis University, Waltham, MA 02454,
USA}\\
\begin{abstract}
In cosmological backgrounds, there can be 'partially massless'
higher spin fields which have  fewer degrees of freedom than their massive 
partners. The equations for the partially massless spin-2 fields are 
usually taken to be the linearized Einstein equations augmented with 
a 'tuned' Pauli-Fierz mass. Here, we add more powers of 
curvatures and show that for 
the string-generated Einstein-Gauss-Bonnet model, partially massless 
spin-2 fields have real mass in AdS, in contrast to 
the Einstein level result. We discuss the implication of this for
the AdS/CFT applications and briefly study the $C^4$-corrected 
$AdS_5\times S^5$ solution in type IIB SUGRA. 
\end{abstract}

\end{center}

\vspace*{1.5cm}

\newpage

\normalsize
\baselineskip=23pt plus 1pt minus 1pt
\parindent=27pt
\vskip 4 cm

In Anti-de-Sitter and de-Sitter spacetimes, 
massive higher spin fields ( $  s \ge 2$ ) exhibit 
a novel phenomenon: partial masslessness (or the appearance of 
extra gauge invariances), if the masses of the fields are `tuned' to the 
background cosmological constant \cite{nepomechie,deser1,deser2}. 
A massless spin-2 field (which we shall 
exclusively employ with here) in a flat $D$-dimensional background 
has $D(D-3)/2$ degrees of freedom ( DOF) and  a massive one has 
$(D+1)(D-2)/2$ DOF. On the other hand, in 
a constant curvature background there is a third possibility, for which 
the spin-2 field has one fewer than the generic massive 
case: Hence the name partially massless. A second derivative,  
scalar gauge invariance appears at the partially massless point.
For these fields, there is a crucial difference between dS 
and the AdS spaces: The tuned $m^2$ is positive for 
the former and negative for the latter.

In the literature, the original analysis involves 
the equations of the linearized gravity 
augmented with the linear Pauli-Fierz mass term.
In this note,  we shall slightly generalize by considering also 
(the linearizations of ) higher curvature terms that 
are generated in certain string 
models and study their effects on the partially massless fields. 

The motivation for this work comes from the possible applications 
of partially massless spin-2 fields to AdS/CFT duality 
\cite{maldacena,polyakov,witten}. As is well-known, bulk fields in AdS
correspond to operators in the boundary conformal field theory. 
According to the AdS/CFT 
dictionary, the masses of the bulk fields determine the 
conformal dimensions of the 
boundary fields. In particular, it was shown by Dolan {\it{et. al.}}
\cite{dolan}, that partially 
massless spin-2 fields $(h_{\mu \nu})$ in the bulk of AdS space 
correspond, on the 
boundary, to  fields $(L_{i j})$, 
which obey a specific conformally-invariant differential 
equation. If the boundary of the AdS is flat, the differential 
operator reduces to partial conservation equation of a tensor field 
(operator) : $\partial_i \partial_j L^{ij}=0$.
 As demonstrated in {\cite{ dolan}}, due to $m^2$ being negative 
for partially massless fields 
in AdS, the first descendant of the corresponding partially 
conserved operator has a negative norm, rendering the boundary 
theory unphysical. 
Therefore, at the linear level there seems to be no room for the 
partially massless fields in AdS. For dS, on the other hand, 
we do not have a proper dS/CFT correspondence to make use of 
these fields. What happens beyond the (already studied ) quadratic
Einstein level, in AdS, is our main aim in this work.

Here we suggest that, in certain cases 
( though, not in the most famous $AdS_5\times S^5$ example), stringy 
higher curvature corrections flip the sign of $m^2$ 
in AdS.  In particular, we will show 
that if the equations of motion for the massive spin-2 field in an AdS 
background are taken to be the linear version of the Einstein plus 
Gauss-Bonnet theory, the partially massless fields have positive mass.

We also study the $({\mbox{Weyl})^4}$ corrected spin-2 action in the 
maximally supersymmetric $AdS_5\times S^5$ case, and show that 
partially massless fields are ruled out here, supporting the conclusion 
of \cite{dolan}. [ Strictly speaking, by imposing the Pauli-Fierz term, 
which does not  arise in string theory, we are changing the 
background geometry. ]

We first review the main ingredients of the partially massless spin-2 
fields in a cosmological background, and then consider the spin-2 field 
equations derived from higher curvature theories.
Finally, we study the $AdS_5 \times S^5$
background as a solution to Type IIB String/SUGRA equations with $R^4$ 
quantum corrections. 

Our conventions are:  signature $(-,+,+, ... +)$,
$[\nabla_\mu, \nabla_\nu]V_{\lambda} =
R_{\mu \nu \lambda}\,^\sigma V_\sigma $,\,\,
$ R_{\mu \nu} \equiv R_{\mu \lambda \nu}\,^\lambda$. We derive 
the equations of motions of a spin 2-field by linearizing 
the gravity part of the low energy string-generated gravity 
equations around the relevant background. 
Let us start with the lowest order Einstein equations 
\be
R_{\mu \nu} - {1 \over 2} g_{\mu \nu} R + \Lambda g_{\mu \nu} = 0,
\label{einstein}
\ee
linearized around $\bar{g}_{\mu \nu}$, for which
the Ricci tensor is $\bar{R}_{\mu \nu}= 
{2 \over {D- 2}}\Lambda \bar{g}_{\mu \nu}$.
We obtain the following covariantly conserved
equation for the massless spin-2 field $h_{\mu \nu} \equiv g_{\mu \nu}- 
\bar{g}_{\mu \nu}$
\be
{\cal{G}}^L_{\mu \nu} \equiv
R_{\mu \nu}^L - {1\over 2} \bar{g}_{\mu \nu} R^L -
{2 \over D-2}\Lambda h_{\mu \nu} =0 ,
\label{lineins}
\ee
where $R^L= (g^{\mu \nu}R_{\mu \nu})_L$ and 
the linear part of the Ricci tensor is defined as
\be
R_{\mu \nu}^L \equiv  R_{\mu \nu} - \bar{R}_{\mu \nu} =
{1\over 2} ( - {\bar{\Box}} h_{\mu\nu} -
{\bar{\nabla}}_\mu {\bar{\nabla}}_\nu h  +
{\bar{\nabla}}^{\sigma}{\bar{\nabla}}_\nu h_{\sigma \mu} +
{\bar{\nabla}}^{\sigma}{\bar{\nabla}}_\mu h_{\sigma \nu} ).
\label{linearricci}
\ee
All derivatives are with respect to the background metric, which also 
raises and lowers the indices. Here: $h = \bar{g}^{\mu \nu}h_{\mu \nu}$ and
$\bar{\Box} = \bar{g}^{\mu \nu}\bar{\nabla}_{\mu} \bar{\nabla}_\nu$.
To get a massive spin-2 field, we add the usual ghost-free 
Pauli-Fierz mass, with the correct relative sign:  
\be
{\cal{G}}^L_{\mu \nu} + {1\over 2} m^2 ( h_{\mu \nu} -\bar{g}_{\mu \nu}h )=0
\label{pauli1}
\ee
For $m^2 \ne 0$, taking the double divergence and the trace of this equation, 
one obtains, respectively:
\be
&& \bar{\nabla}_{\mu} \bar{\nabla}_{\nu} h^{\mu \nu} - \Box h  =0, \\
&&  \left [\Lambda - {(D-1)\over 2}m^2 \right ] h =0.
\ee
Unlike the generic $(m^2, \Lambda)$ theory, $h_{\mu \nu}$ does not have to 
be traceless ({\it{i.e}} $h \ne 0$) at the partially massless point for which
the mass is tuned as 
\be
m^2 = {2\over D-1}\Lambda
\label{massequation}.
\ee
One condition on the field drops out, yielding instead the following 
(higher derivative) scalar gauge invariance
\be
\delta h_{\mu \nu} = \bar{\nabla}_\mu \bar{\nabla}_\nu \xi + 
{2\over (D-1)(D-2)}\bar{g}_{\mu \nu}\Lambda \xi
\ee
Needless to say (\ref{massequation}) assigns a negative $m^2$ for 
the partially massless fields in AdS. This by itself does not pose a threat 
since AdS allows negative $m^2$ for various fields, including massive 
spin-2 field as long as the corresponding Brietenlohner-Freedman 
\cite{freedman} bounds are  satisfied.

In \cite{dolan}, whose notations and results we follow, 
partially massless spin-2 field was employed 
in the context of AdS/CFT. The authors showed that such a field 
in the bulk of $AdS_{D}$ corresponds to a symmetric tensor 
$L_{ij}$ on the boundary, which satisfies the following 
conformally invariant differential equation
\be
\left [ \bar{\nabla}_i \bar{\nabla}_j +{1\over D-3} \tilde{R}_{ij} \right ] 
L^{ij} =0, 
\label{condiff}
\ee  
where  ${\tilde{g}}_{ij}$ is the conformal metric on the boundary.
This differential equation was studied before by Eastwood and collaborators
\cite{eastwood}. As in the simplest examples of AdS/CFT, 
let us assume that the boundary is conformally flat, then (\ref{condiff})
reduces to the more transparent partial conservation law
\be
\partial_i \partial_j L^{i j} =0.
\label{flatcase}
\ee 
Suppose now $|L^{ij} >$ is a symmetric primary state on which the $D-1$ 
dimensional conformal group acts. Let $P_i$ be the momenta; as a  
a result of (\ref{flatcase}), we require $P_i P_j | L^{ij} >$
to be a null state. This then implies the existence of a 
negative norm state: $ ||P_i L^{ij}>|^2 < 0$. In the language of the 
bulk fields, this negative norm state 
corresponds to $m^2$ being negative for the partially massless AdS fields 
(\ref{massequation}). 

At the linear level, the above result signals an impasse 
for the partially massless fields in the context of AdS/CFT. 
But here, I will argue that higher curvature corrections 
can drastically change the linear level results in certain 
theories. In fact, though rare, it is not unheard-of  
that higher derivative terms alter the sign of the linear level theory.
[ To warm up, let us recall the non-trivial example of the 
Topologically Massive Gravity 
(TMG) in $2+1$ dimensions \cite{jackiw}. This model is 
constructed by adding a third derivative Cotton tensor (or 
Chern-Simons term ) to the Einstein one $
R_{\mu \nu} -{1\over 2}g_{\mu \nu}R + {1\over M} C_{\mu \nu} =0.$
Unlike pure 3D Einstein gravity, TMG is a dynamical theory: 
Linearized equations show that there is a massive scalar degree of 
freedom with mass $|M|$. For this 
scalar mode to be non-ghost with positive 
Hamiltonian, the {\it{sign}} of the Einstein  {\it{action}} 
has to be flipped from its usual one once the higher derivative 
topological term is introduced \cite{jackiw}. 
Even though TMG is quite distinct from the 
theories we are about to consider, it nevertheless
provides us with a concrete example of how higher derivative terms 
can change linear level signs.]

Let us now turn to Einstein-Gauss-Bonnet model \cite{zweibach}, 
which is expected to arise in some string theories [ or string 
theory compactifications].
\be
I = \int d^D\, x \sqrt{-g} \Big \{ {R\over \kappa} +
\gamma (R_{\mu\nu\rho\sigma}^2 -4R^2_{\mu\nu}+R^2 ) \Big \}.
\label{action}
\ee
String theory dictates the sign of $\gamma$ to be positive.  
Even without an explicit cosmological constant, AdS [not dS!] is a solution
\cite{boulware} to the equations of motion derived from (\ref{action})
\be
&& 0= {1\over \kappa} (R_{\mu \nu} -{1\over 2}g_{\mu \nu} R) \nonumber \\
&&+ 2\gamma \Big \{ R R_{\mu \nu} - 2 R_{\mu\sigma\nu\rho}R^{\sigma \rho}
+ R_{\mu\sigma\rho\tau}R_\nu^{\sigma\rho\tau} - 2R_{\mu \sigma}R_\nu^\sigma
- {1\over 4}g_{\mu \nu}(R_{\tau\lambda\rho\sigma}^2 -4R^2_{\sigma\rho}+R^2 )
\Big \}. 
\label{eom}
\ee
The cosmological constant of the AdS space that solves the equations
is determined by ($\gamma, \kappa$)
\be
\Lambda = -{2(D-1)(D-2)\over \kappa \gamma (D-3)(D-4)} 
\label{lam1}.
\ee
Linearizing (\ref{eom}) about AdS, one finds \cite{tekin} how  
the presence of $\gamma$ modifies the massive spin-2 equation,
\be
{\cal{G}}^L_{\mu \nu} \left [{1\over \kappa} + {4\Lambda \gamma(D-4)(D-3)\over 
(D-2)(D-1)}\right ]
 + {1\over 2 \kappa } m^2 ( h_{\mu \nu} -\bar{g}_{\mu \nu}h )=0
\label{pauli2}
\ee
subject to the condition (\ref{lam1}). Thus one actually obtains the
{\it{sign-flipped}} version of (\ref{pauli1})
\be
{\cal{G}}^L_{\mu \nu} - {1\over 2} m^2 ( h_{\mu \nu} -\bar{g}_{\mu \nu}h )=0
\label{pauli3}
\ee
The crucial and non-trivial point here is that 
both (\ref{pauli1}) and (\ref{pauli2}) 
come from proper, non-ghost Lagrangians, despite `contradicting' each other.
It is , by now, clear that the partially massless states  in AdS
will be non-tachyonic for Einstein-Gauss-Bonnet theory. Namely for 
(\ref{pauli3}): $m^2 
= - 2 \Lambda/(D-1) > 0 $. It also follows that if 
this model describes spin-2 fields in the bulk 
of $AdS_D$, for $D \ge 5$, partially massless fields correspond 
to conformally invariant boundary operators with positive descendants. 
Thus adding the Gauss-Bonnet term would circumvent the objection 
of \cite{dolan} ( at least beyond four dimensions ). 

On the other hand, if we include an explicit cosmological constant (
$\Lambda_0$, whose sign is arbitrary ), the above 
analysis branches into several 
directions. For given 
$(\Lambda_0, \kappa, \gamma)$ there are two different AdS spacetimes which 
solve the equations. [ See the related work \cite{padilla}]. 
The cosmological constant of these spaces 
are
\be
\Lambda_{\pm} =  {-1 \pm \sqrt{ 1 + 8\kappa \tilde{\gamma}\Lambda_0} \over
4\kappa \tilde{\gamma}}
\ee
where $\tilde{\gamma} = \gamma (D-4)(D-3)/[(D-2)(D-1)]$ and  
$8\kappa \Lambda_0 \tilde{\gamma} \geq -1$.  
The equation (\ref{pauli3}) becomes
\be
{\cal{G}}^L_{\mu \nu} \pm {1\over 2} {m^2\over 
\sqrt{ 1+ 8\kappa \Lambda_0 \tilde{\gamma}}}
 ( h_{\mu \nu} -\bar{g}_{\mu \nu}h )=0
\ee
and the partially massless fields have the mass
\be
m^2 = \pm {2 \Lambda_{\pm}\over D-1}\sqrt{ 1+ 8\kappa \Lambda_0 \tilde{\gamma}}
\ee
One, now, has various choices, depending on the sign of $\Lambda_0$ and 
on the choice of $\Lambda_{+}$ or $\Lambda_{-}$. For $\Lambda_0 <0$, we have 
$\Lambda_{\pm} <0$, but only $\Lambda_{-}$ gives real mass to partially 
massless fields.

Having pointed out that certain higher curvature spin-2 theories have 
non-tachyonic masses for the partially massless fields, can we now make use 
of this result in AdS/CFT? Unfortunately, I do not know any examples of 
latter where there are Gauss-Bonnet corrections to bulk gravity.
One would expect that less supersymmetric forms of AdS/CFT duality
might allow Gauss-Bonnet terms [See \cite{gregory} and the references 
therein.] But, for now, let us briefly study
the well-understood maximally supersymmetric case.
In low energy Type IIB SUGRA the next to leading order corrections 
(in the $\alpha'$ expansion ) around the         
$AdS_5\times S^5$ vacuum are of the $C^4$ type( here $C^4$ 
schematically represents the two inequivalent contractions of four Weyl 
tensors $C_{\mu\nu\alpha \beta}$ 
\cite{banks}. Switching off the dilaton and all the fields but spin-2, the 
action reads
\be
I = {1 \over {\alpha'}^4} \int d^{10}x \sqrt{-g} \left ( R + 
k{\alpha'}^3 f_4(\rho,\bar{\rho})C^4 \right )   
\label{type2b}
\ee
where $\rho$ is the usual complex coupling 
and  $f_4$  is given by an Eisenstein series, whose details
will not be relevant to us, but  
can be found in \cite{banks}. $k$ is a constant. $AdS_5\times S^5$
is a background solution. ( Due to the flux from the
self-dual five form $F_5$, a cosmological constant is generated.)
$AdS_5\times S^5$, with the same radius on $AdS_5$ and 
$S^5$ is conformally flat. This vacuum is 
expected to be a solution to full string theory:Thus the appearance of $C^4$,
which vanishes for the conformally flat geometries, is no surprise.
To see if the higher order terms can affect the partially massless 
fields, one can linearize the equations derived from (\ref{type2b}) 
around $AdS_5\times S^5$ and add a Pauli-Fierz mass term. But it is 
immediately clear that there will be no contributions from the $C^4$ 
terms since, at $O(h^2)$ level one has 
$\bar{C}^2 \delta(\sqrt{-g}C^2) $, which vanishes for all
conformally flat spaces. Thus $C^4$ corrections cannot change 
the structure of the partially massless fields. With 
negative $m^2$, since the latter predict the existence of negative norm 
states in the boundary ${\cal{N}} = 4$ $SU(N)$ Yang-Mills 
theory, they ought to be ruled out in the bulk. It is important 
to note that if the corrections were of the generic $R^n$-type, 
where $R$ represents scalar,Ricci, or Riemann curvatures, the 
equations for the spin-2 field and its partially massless point would 
have changed drastically. On the other hand $C^3$, and $C^2$ corrections
do not change the picture nor do arbitrary power of $C$. 
Up to $O(h^2)$, $C^3$ terms vanish just like the $C^4$ ones
for AdS. Even though $C^2$ does not vanish at this order, it does 
not contribute to the trace of the equations, and thus  does not affect 
the partially massless point.

To summarize, we have shown that, unlike the linear spin-2 theory,  
partially massless fields have {\it{positive}} $m^2$ in AdS, if 
Gauss-Bonnet corrections are added for $D \ge 5$. In the context of 
AdS/CFT, these partially massless fields in the bulk correspond to
operators with positive norm descendants in the boundary.

This paper is dedicated to the memory of Ian Kogan (1958-2003). 
I would like to thank S. Deser and A. Waldron for useful discussions.
I also would like to thank A. Lawrence, H. Schnitzer and N. Wyllard 
for directing me to the references regarding the $R^4$ corrections in 
string theory. This research was supported by NSF Grant 99-73935.

\myend